# Sensorless Field Oriented Control of CSI-Fed PMSM Drives Used in Submersible Pumps

Milad Bahrami-Fard, *Student Member, IEEE*, Majid Ghasemi Korrani, *Student Member, IEEE*, and Babak Fahimi, *Fellow, IEEE*
Renewable Energy and Vehicular Technology (REVT) lab
Department of Electrical Engineering, University of Texas at Dallas
Richardson, TX, USA
Email: milad.bahramifard@utdallas.edu

*Abstract*—This paper proposes a practical startup strategy for current source inverter (CSI)-fed Permanent Magnet Synchronous Motor (PMSM) drives in submersible pump applications, focusing on ensuring a seamless shift to sensorless field-oriented control (FOC). The method effectively manages the transition to sensorless operation without requiring precise current or alignment error calculations, thereby simplifying implementation. By addressing speed and current oscillations directly during the startup and transition stages, the approach significantly enhances overall system stability and responsiveness. Validation through simulation and experimental testing demonstrates the strategy's success in maintaining low oscillation levels across various operating conditions, confirming its reliability for high-performance industrial applications.

## I. INTRODUCTION

In modern industrial applications, there is a continuous push for innovations that improve efficiency, precision, and operational reliability. Recent advancements in control strategies and power electronics have increasingly focused on enhancing the reliability and adaptability of Permanent Magnet Synchronous Motor (PMSM) drives in demanding industrial environments. PMSM drives are considered in industrial applications for their high efficiency, dynamic performance, and power density, making them ideal for tasks demanding precise speed and torque control [1]-[7]. To achieve these attributes, Field-Oriented Control (FOC) has become prevalent, as it decouples motor flux and torque components, enabling accurate control. Traditionally, high-performance FOC strategies rely on position sensors that provide accurate rotor position and speed measurements, optimally positioning the stator magnetic field with the rotor flux for optimal performance [8]. However, position sensors introduce higher costs and reduced reliability, particularly under extreme environmental conditions, such as submersible pumps that can operate several kilometers below the surface, where installing and maintaining sensors becomes extremely challenging [9].

Position sensorless control algorithms have therefore been developed to eliminate the need for physical sensors. These sensorless methods are categorized broadly into high-frequency signal injection (HFI) and model-based observer techniques. While HFI techniques excel at detecting rotor position at low or zero speeds by analyzing high-frequency current responses due to rotor magnetic saliency, they are highly sensitive to electrical noise and require complex filtering, complicating real-world implementation [10]-[13]. In contrast, model-based methods—better suited for medium to high speeds—estimate back electromotive force (EMF) and utilize a phase-locked loop (PLL) to track rotor position and velocity [14]-[17]. For comprehensive sensorless FOC across the full speed range, these methods are often combined. However, this dual-method approach increases controller complexity and can reduce reliability, particularly in demanding applications like submersible pumps, where simplicity and robustness are essential [18]. Furthermore, in submersible pump application, access to voltage and current at motor terminal is not trivial and injection of high frequency field components will face similar challenges across a few kilometers of cables that connect the inverter to the motor.

Current Source Inverters (CSIs) bring additional benefits to PMSM drives, particularly in industrial applications requiring durability and robustness. CSIs support four-quadrant operation, short-circuit protection, and voltage-boosting capability while providing motor-friendly waveforms with low dv/dt, all of which enhance the connected PMSM's reliability and longevity [19]-[21]. This makes CSI-fed drives well-suited for high-power, medium-voltage applications, especially in industries like petrochemicals, mining, and metal processing, as well as in pumping applications requiring long cable runs, where they outperform voltage source inverters (VSIs) that struggle with capacitive cable effects [22]. Although sensorless control research has largely focused on VSI-fed drives, studies on sensorless control for CSI-fed PMSMs are still relatively limited and call for more attention, keeping in mind that access to accurate motor currents in CSI is a key element for successful implementation of sensorless control.

To perform sensorless control at medium to high speeds, methods based on back EMF estimation are particularly effective, provided that there is a reliable startup strategy to reach a speed where the back EMF can be reliably sensed. Conventional startup methods like V/f control, common in induction motors, are simple and open-loop but lack closed-loop current feedback, leading to instability under heavy loads [23]. The I-f method, first implemented in PMSMs in [24], offers an alternative with closed-loop current regulation and does not require position feedback. It approximates rotor position using a ramped speed command while maintaining constant reference-frame current, delivering a smoother startup and reduced torque ripples compared to V/f [25]. However, the I-f method can result in a misalignment between the synchronous and actual rotor frames, especially under high startup currents, as seen in submersible pump applications [26]. Accurate frame alignment is crucial for a successful transition from I-f startup to sensorless FOC, as misalignment introduces rotor position estimation errors. This, in turn, challenges the transition and potentially destabilizes the dynamics of speed [27]-[31].

Several methods have been proposed to manage this transition. For example, [24] used a first-order compensator to align the synchronous reference frame with the rotor position, facilitating a smooth shift from I-f control to EMF-based control, though the effectiveness depended heavily on compensator tuning. In [32], a gradual decrease in angle gradient deviation was achieved by adjusting it based on the reference current as the motor reached the target speed, allowing a smoother transition but extending the startup duration. Alternatively, [33] applied feedforward compensation within the current loop, enhancing stability but primarily prolonging the I-f starting period. Lastly, [34] proposed a speed-dependent weighting factor to align the synchronous reference frame to the rotor position, yet finding the correct parameters for convergence remained challenging, and no detailed stability analysis for I-f methods was provided.

In this paper, a new startup strategy is introduced using which a smooth transition from I-f startup to sensorless FOC for CSI-fed PMSM drives is obtained. Unlike previous methods that depend on precise frame alignment or tuning the $q$-axis current, our approach uses an error compensation strategy to eliminate frame alignment errors. By first aligning the estimated frame to a virtual reference and then applying gradual error correction, stable speed and current profiles have been achieved without requiring complex parameter calculations. This strategy allows for accurate rotor position estimation without calculating the minimum $q$-axis current or other parameters.

The rest of this article is structured as follows: Section II reviews sensorless control fundamentals for CSI-fed PMSM drives, including EMF-based observer derivation, and provides a detailed analysis of the I-f startup process, examining primary non-ideal factors impacting rotor position estimation along with the proposed transition strategy. Section III presents simulation results validating the method, followed by experimental confirmation in Section IV. Section V concludes with a summary of contributions and potential future directions.

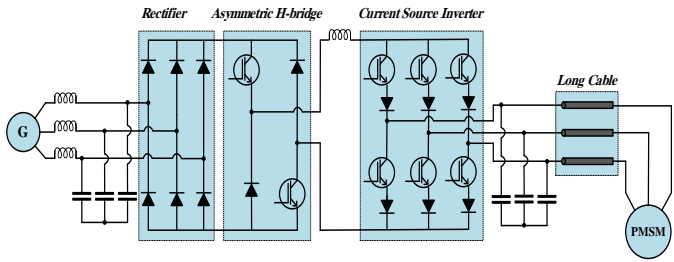

Fig. 1. CSI-fed PMSM drive system with a long cable connection for submersible pump applications.

## II. SENSORLESS CONTROL OF THE CSI-FED PMSM WITH AN IMPROVED STARTUP

Fig. 1 depicts the CSI-fed PMSM drive system with a long cable connection, comprising a passive rectifier, an asymmetric H-bridge, a current source inverter, a DC inductor, and output capacitor filters. The asymmetric H-bridge regulates the DC current to maintain a constant DC-link current. The motor speed and torque are controlled by a CSI using a FOC.

### A. Vector Control of Sensorless CSI Fed PMSM

The block diagram of the sensorless FOC for the CSI-fed PMSM based on the back-EMF observer is illustrated in Figs. 2 and 3. The control sets the $d$-axis current to zero while regulating the $q$-axis current according to the set value of the speed controller.

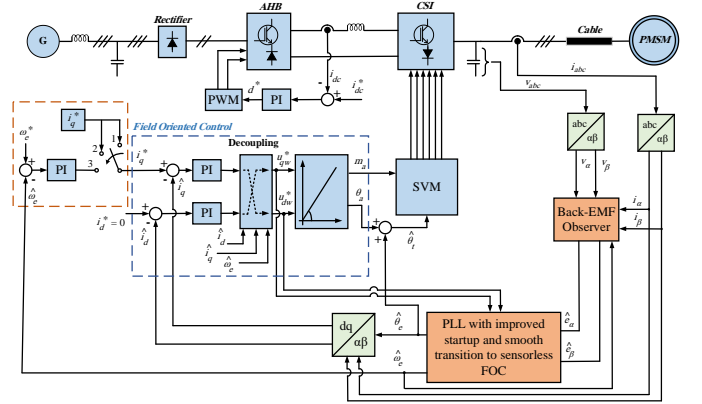

Fig. 2. FOC control of the CSI-fed PMSM drive system with proposed smooth transition strategy

### B. Mathematical Model of the PMSM

To start the process of the position estimation for PMSMs, the mathematical model of the surface-mounted PMSM, expressed in a stationary reference frame (αβ), is given as follows.

$$L_s\left(\frac{di_\alpha}{dt}\right) = -R_s i_\alpha - e_\alpha + u_\alpha \quad (1)$$

$$L_s\left(\frac{di_\beta}{dt}\right) = -R_s i_\beta - e_\beta + u_\beta \quad (2)$$

where $i_\alpha$, $i_\beta$ and $u_\alpha$, $u_\beta$ are the stator currents and voltages, respectively, $e_\alpha$ and $e_\beta$ are the back EMFs, $R_s$ is the stator resistance, and $L_s$ is the synchronous inductance. The back EMF for each phase can be described as

$$e_\alpha = -\frac{\sqrt{3}}{2}\psi_f P\omega_e \sin(\theta_e) \quad (3)$$

$$e_\beta = \frac{\sqrt{3}}{2}\psi_f P\omega_e \cos(\theta_e) \quad (4)$$

where $\psi_f$ is the flux linkage due to rotor permanent magnet, $P$ is the number of pole pairs, $\omega_e$ is the rotor angular speed, and

$\theta_e$ is the actual rotor position. The α- and β-axis back EMF inherently contain the actual rotor position $\theta_e$, but direct measurement of these values is not feasible. Therefore, an observer along with a PLL will be utilized to derive the position information.

To estimate the position of PMSMs, mathematical model of a surface-mounted PMSM, expressed in $dq$-frame, is given as follows.

$$L_s\left(\frac{di_d}{dt}\right) = -R_s i_d + \omega_e L_s i_q + u_d \quad (5)$$

$$L_s\left(\frac{di_q}{dt}\right) = -R_s i_q - \omega_e L_s i_d - \omega_e \psi_f + u_q \quad (6)$$

where, $i_d$, $i_q$ and $u_d$, $u_q$ are the stator currents and voltages, respectively, $\omega_e$ is the rotor electrical speed, $\psi_f$ is the PM-originated flux linkage, $R_s$ is the stator resistance, and $L_s$ is the synchronous inductance.

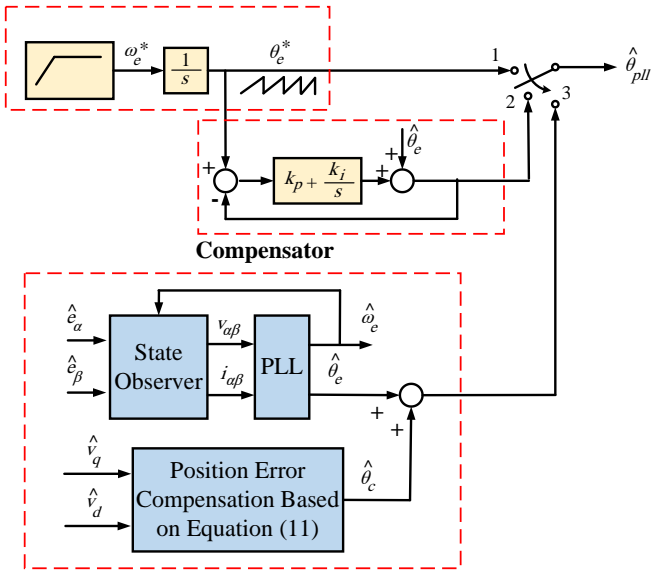

Fig. 3. Diagram of the PLL with improved startup and smooth transition to sensorless FOC

*C. I-f Startup Analysis*

During the speed ramp-up, the startup procedure uses a closed current loop and an open speed loop in the synchronous reference $d^*q^*$-frame, as shown in Figs. 4 and 5. In this process, the reference $q^*$-axis current ($i_q^*$) is maintained constant, and the $i_d^*$ is set to zero. The reference $\theta_e^*$ is obtained by integrating the ramp speed command ($\omega_e^*$) as follows:

$$\theta_e^* = \int \omega_e^* dt \quad (7)$$

$$\omega_e^* = K_\omega t \quad (8)$$

where $K_\omega$ is a constant. At the start of the procedure, the $d^*q^*$-frame is set to a $\delta_e^* = 90°$ phase lag relative to the real rotor $dq$-frame so that the $dq$-frame can start moving smoothly, as shown in Fig. 4. The $d^*q^*$-frame rotates as the ramp speed command is integrated. As angle $\delta_e^*$ decreases, the $q$-axis current ($i_q$) increases proportional to $i_q^* \cos \delta_e^*$. Once $i_q$ generates sufficient torque to exceed the load torque, the rotor begins to rotate, and $\delta_e^*$ converges to a fixed value depending on the load and acceleration. The relationship between the $d^*q^*$-axis currents and the actual currents is defined as follows:

$$i_q = i_q^* \cos \delta_e^* - i_d^* \sin \delta_e^* \quad (9)$$

$$i_d = i_q^* \sin \delta_e^* + i_d^* \cos \delta_e^* \quad (10)$$

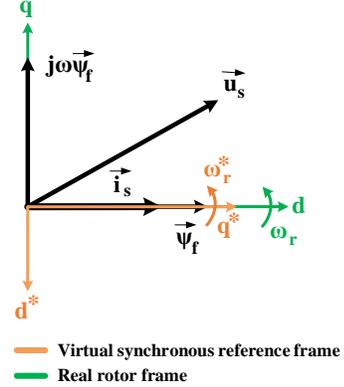

Fig. 4. Diagram of the virtual synchronous $d^*q^*$-reference frame and real rotor $dq$ frame at initial startup

If angle $\delta_e^*$ becomes negative, the motor's self-stabilization is lost, leading to a rapid decrease in speed and potentially stall of the rotor. To prevent this, it is crucial to keep the $d^*q^*$-axis lagging behind the rotor $dq$-axis with sufficient margin in angle $\delta_e^*$. This is typically achieved by maintaining a high enough $q^*$-axis current. However, when $i_q^*$ is greater than $i_q$, it creates an angle error ($\delta_e^*$) between the synchronous reference frame and the real rotor frame. Therefore, an immediate switching of the speed controller from open-loop startup to closed-loop FOC control mode can cause large torque and current pulses.

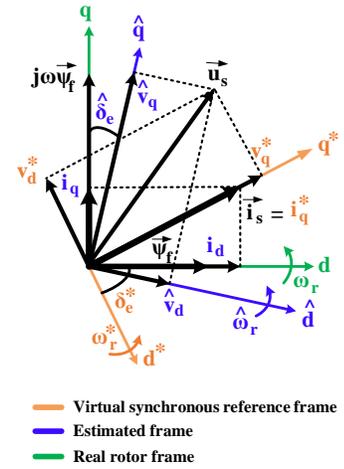

Fig. 5. Diagram of the virtual synchronous $d^*q^*$-reference frame, real rotor $dq$ frame, and estimated $\hat{d}\hat{q}$ frame during acceleration

The estimated rotor axis is denoted by the $\hat{d}\hat{q}$-axis, and Fig. 5 illustrates its relationship to the $dq$-axis. The estimated position is shown by $\hat{\theta}_e$, and the actual position is denoted by $\theta_e$. As previously discussed, performance of FOC is directly influenced by the precision of rotor position estimation. There are multiple non-ideal elements that contribute to position estimation error. These include, saturation-related deviations in $L_s$, computation time, feedback-related control delays in the motor drive system, and Pulse Width Modulation (PWM) of the inverter, all of which lead to errors proportionate to rotor speed. It is difficult to quantify these inaccuracies, especially those caused by control delay ($T_d$). The impact of iron loss is also highly significant at high speeds, however its quantification is challenging since it depends on multiple non-ideal criterions. Furthermore, because of the substantial deadtime ratio in a single switching period at high speeds, the inverter's insertion of deadtime needs a greater command voltage than the actual voltage, leading to severe position estimate inaccuracies.

Defining the difference between the estimated and the real positions as position error ($\hat{\delta}_e = \theta_e - \hat{\theta}_e$), the equations (4) and (5) can be transformed into the following expressions in $\hat{d}\hat{q}$-axis under steady state condition.

$$\hat{u}_q = \omega_e L_s \hat{i}_d + \omega_e \psi_f \cos(\hat{\delta}_e) \quad (11)$$

$$\hat{u}_d = -\omega_e L_s \hat{i}_q + \omega_e \psi_f \sin(\hat{\delta}_e) \quad (12)$$

The difficulty in accurately measuring the entire estimation error caused by many non-ideal elements makes it difficult to directly compensate for $\hat{\delta}_e$. Therefore, it is necessary to engineer a method that can compensate position errors $\delta_e^*$ and $\hat{\delta}_e$ before transitioning from open-loop startup to closed-loop FOC control mode.

TABLE I. SYSTEM PARAMETERS

| Parameters | Value |
|---|---|
| equivalent resistance $R_s$ | 2.16 Ω |
| equivalent inductance $L_s$ | 4.56 $mH$ |
| cable equivalent resistance $R_c$ | 11.76 Ω |
| cable equivalent inductance $L_c$ | 9.7 $mH$ |
| cable equivalent capacitance $C_c$ | 111 nF |
| number of pole pairs $P$ | 6 |
| grid voltage $V_g$ | 480 V |
| grid frequency $f_g$ | 60 Hz |
| DC-link inductance $L_{dc}$ | 10 mH |
| output filter $C_o$ | 50 uF |

### D. Proposed Smooth Transition to Senesorless Control Strategy

The block diagram in Figs. 2 and 3 illustrates a sensorless FOC system for a CSI-fed PMSM, designed with a new approach for smooth transition to sensorless control. This system is implemented through three consecutive steps, each represented by a distinct reference frame configuration across the startup and transition phases.

In the initial step, the I-f method activates the motor from a standstill with controlled acceleration to reach a stable target speed. This provides sufficient back EMF for position estimation using a PLL and observer, with terminal 1 selected. During this period, the $q^*$-axis current remains constant while the $d^*$-axis current is zero, setting up the synchronous $d^*q^*$-reference frame to lag with respect to the actual rotor frame by 90 degrees. As the synchronous frame begins its rotation, the real $q$-axis current ($i_q$) increases, and an angle offset ($\delta_e^*$) emerges between the frames. Once torque from $i_q$ overcomes the load, rotor rotation aligns the rotor and synchronous frames.

Next, to minimize the position offset between the estimated and the actual reference frames, a compensator brings the error to zero, preparing the system for terminal 2. Here, only the position information will update while $i_q^*$ remains fixed. The alignment of the estimated $\hat{d}\hat{q}$-frame with the virtual synchronous $d^*q^*$-reference frame enables a seamless terminal switch, avoiding any fluctuations in motor speed or current.

In the final step, transitioning to terminal 3 introduces an initial position error from the compensator. This error is systematically reduced using an error correction strategy outlined in equations (11) and (12). This shift maintains continuity in speed and current due to stable outputs from the speed PI controller and position compensation. As a result, the position error diminishes gradually, achieving smooth operation after switching terminals.

Based on (4), to completely correct the position error after shifting to terminal 3, $u_q'$ must be at its maximum value (i.e., $\cos(\hat{\delta}_e) = 1$). The estimated position is adjusted using the proposed method to maintain $u_q'$ at its maximum and to make sure that the position error is compensated for. '$\theta_c$' is designated as a compensation factor utilized to rectify the position error estimation unit. The algorithm starts by initializing $\theta_c$ and assuming a variable step size '$d\theta$'. By setting 'h' as the number of control cycles amplitudes of $u_q'$ and $u_d'$ are calculated. If $u_q'$ increases, $d\theta[h]$ remains constant. Otherwise, the sign of $d\theta[h]$ is altered (i.e., $-d\theta$) and the magnitude of the voltages are recalculated. This method will guarantee the position error will be compensated in a relatively short time without requiring complex models.

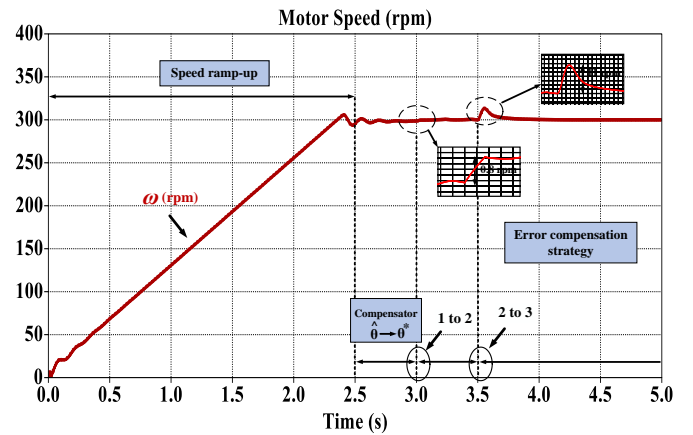

Fig. 6. Simulation waveform of the motor speed using the proposed startup strategy with a smooth transition to sensorless FOC control

## III. SIMULATION RESULTS

The proposed startup strategy for the CSI-fed PMSM drive has been validated through simulation studies. Table I outlines the key parameters of the CSI-fed PMSM drive, while Fig. 6 presents the simulation results, illustrating a seamless transition to sensorless FOC. Using the I-f method, the motor successfully starts from standstill and accelerates to a stable speed of 300 rpm, where the back EMF becomes sufficiently large for position estimation via the PLL and observer. At t=2.5s, the compensator aligns the estimated frame with the virtual reference frame. At t=3s, the terminal transitions from 1 to 2 without inducing any tangible speed or current oscillations. Finally, at t=3.5s, the terminal shifts from 2 to 3, utilizing the proposed compensation strategy from Equation (11) to align the estimated frame with the real frame. The results indicate minimal speed oscillation of about 11 rpm, demonstrating the effectiveness of the smooth transition strategy.

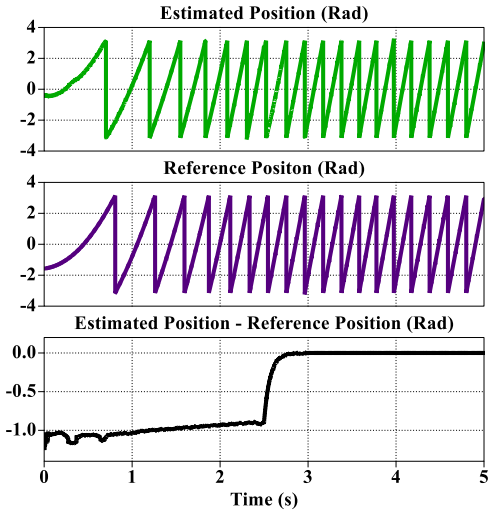

Fig. 7. Simulation results of the proposed startup strategy; Estimated position, Virtual synchronous reference position, Angle error between estimated $\hat{d}\hat{q}$ frame and virtual synchronous $d^*q^*$-reference frame.

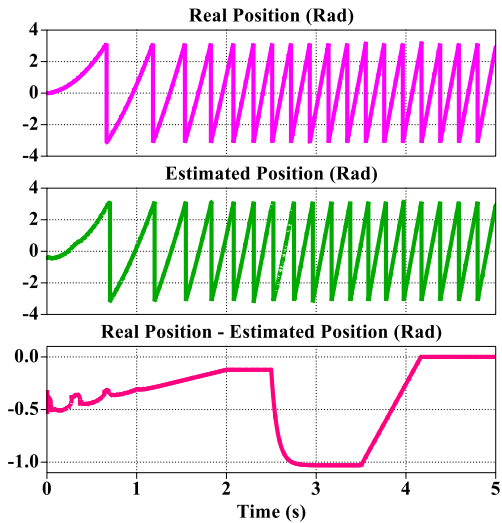

Fig. 8. Simulation results of the proposed startup strategy; Real position, Estimated position, Angle error between actual rotor $dq$ frame and estimated $\hat{d}\hat{q}$ frame

In Fig. 7, the estimated position, virtual synchronous position, and the error between them are plotted. At t=2.5s, the compensator aligns the estimated frame with the virtual reference frame. Once alignment is achieved, the terminal switches from 1 to 2 at t=3s without introducing transients in position, current, or speed. Fig. 8 illustrates the real and estimated positions along with the error between them. As shown, the error between the estimated and real positions is eliminated using the proposed compensation strategy.

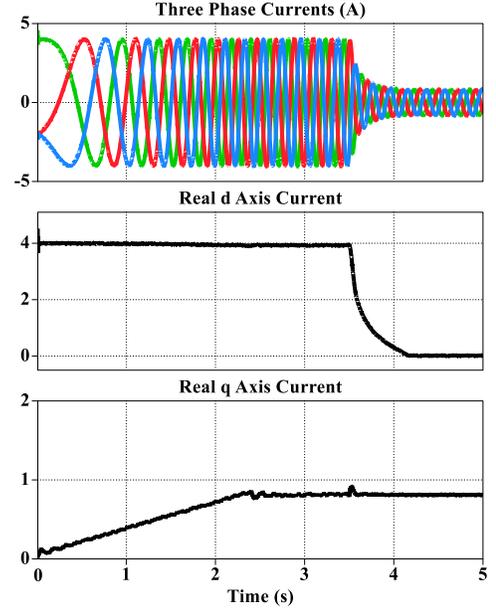

Fig. 9. Simulation results of the proposed startup strategy (a) Three phase currents. (b) Real $dq$-axis currents

Figs. 9 and 10 present the simulation waveforms of the three-phase, $dq$-axis, and $\hat{d}\hat{q}$-axis currents during various stages of the proposed startup strategy. As shown in Fig. 9, the transition from terminal 1 to 2 occurs without any oscillation in the $dq$-axis currents. Furthermore, during the transition from terminal 2 to 3, when the proposed compensation strategy is applied, the $dq$-axis and $\hat{d}\hat{q}$-axis currents converge and become identical after 0.7 s, with minimal oscillation in current (0.09 A).

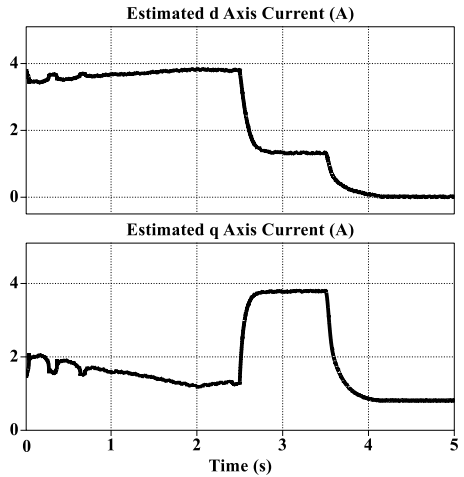

Fig. 10. Simulation results of the proposed startup strategy; Estimated $\hat{d}\hat{q}$-axis currents.

## IV. EXPERIMENTAL RESULTS

To experimentally validate the proposed startup strategy, tests were performed on a CSI-fed PMSM drive system with a 1.8 km long cable connection, using a TMS320F28388D processor, as shown in Fig. 11. An optical encoder was employed exclusively to obtain the actual position for comparison purposes.

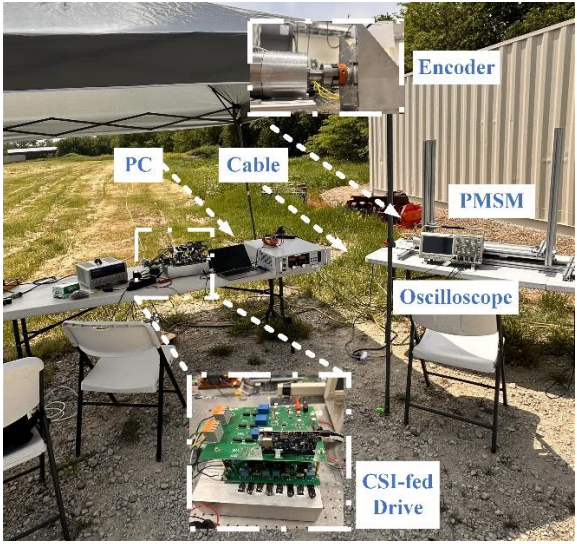

Fig. 11. Experimental setup of the CSI-fed PMSM drive system with a long cable connection for submersible pump.

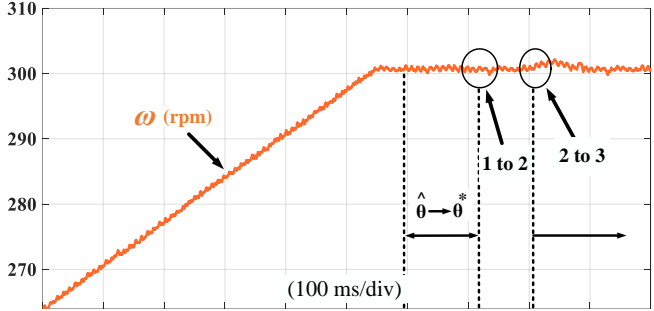

Fig. 12. Experimental results of the motor speed during different stages of the proposed startup strategy.

Fig. 12 shows the experimental results of motor speed during the entire transition from the I-f method to sensorless FOC, with the application of the proposed smooth transition strategy. Similar to the simulation tests, the motor accelerates to 300 rpm, where the estimated speed is accurately captured. The compensator then aligns the estimated position with the virtual reference frame. Following this, the terminal transitions from 1 to 2 without causing any speed or current oscillations. Finally, the transition from terminal 2 to 3 occurs, and the position error between the estimated and actual position is corrected using the proposed strategy. The experimental results validate the effectiveness of the proposed startup method, showing only minimal speed oscillations.

Figs. 13 and 14 present the experimental results of the proposed startup strategy throughout all stages of the transition from the I-f method to sensorless FOC. Initially, the I-f method is used to start the motor from standstill, gradually accelerating to a stable speed where the back EMF becomes high enough to estimate the position using the PLL and observer. At this point, both switches are connected to terminal 1. The compensator reduces the error between the estimated and virtual synchronous positions to zero, preparing the system for the transition to terminal 2. As shown in Fig. 13(a) and (b), the virtual synchronous position and the estimated position are aligned, and the error angle between them is minimized, as demonstrated in Fig. 13(c). After switching to terminal 3, the position error between the estimated and virtual synchronous positions is corrected using the proposed compensation strategy, as described in the equations. The angle error between the estimated and virtual synchronous frames is effectively reduced, demonstrating the success of the error compensation strategy. Additionally, as the system transitions from terminal 2 to 3, the angle error between the actual rotor frame and the estimated frame is eliminated, as shown in Fig. 14(c). The alignment between the estimated and real positions is achieved, as shown in Fig. 14(a) and (b). This confirms the effectiveness of the compensation strategy, with the estimated frame aligning with the real rotor position and achieving minimal error. This alignment is maintained throughout the transition, showcasing the robustness and accuracy of the proposed method.

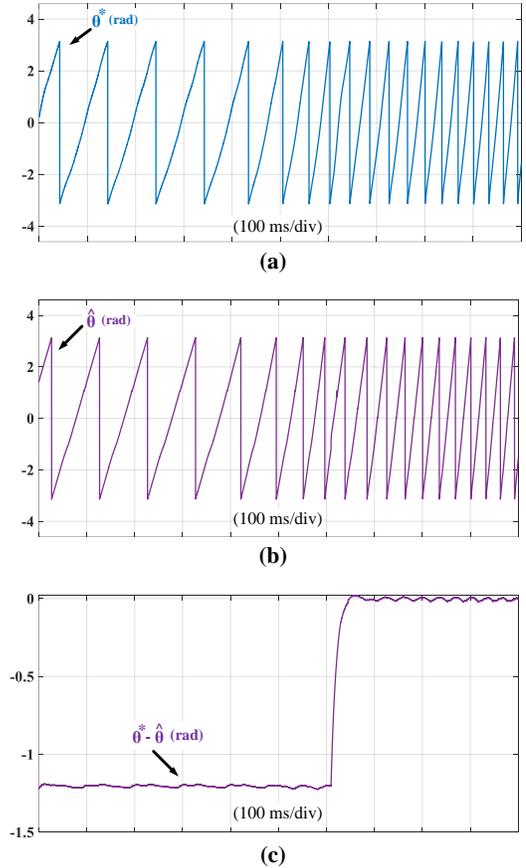

Fig. 13. Experimental results of the proposed startup strategy throughout all stages of the transition from the I-f method to sensorless FOC (a) Virtual synchronous position. (b) Estimated position. (c) Angle error between estimated $\hat{d}\hat{q}$ frame and virtual synchronous $d^*q^*$-reference frame.

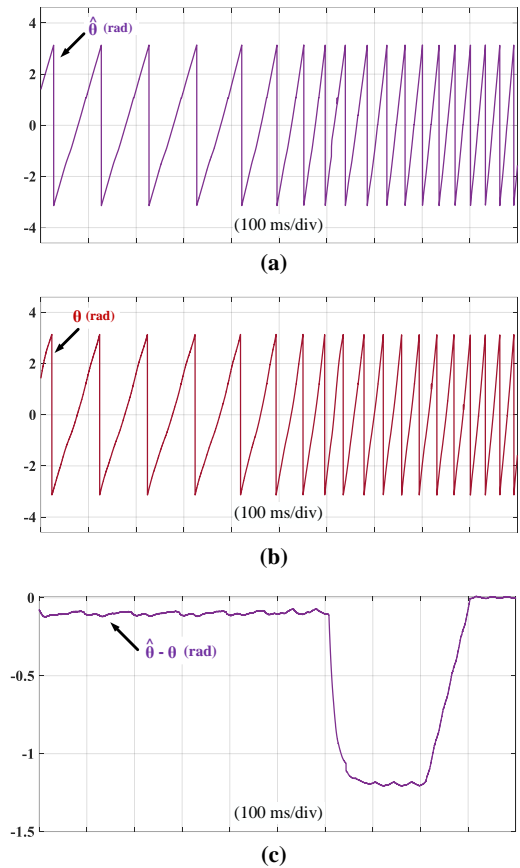

Fig. 14. Experimental results of the proposed startup strategy throughout all stages of the transition from the I-f method to sensorless FOC (a) Estimated position. (b) Real position. (c) Angle error between actual rotor $dq$ frame and estimated $\hat{d}\hat{q}$ frame.

## V. CONCLUSION

This paper introduces a novel startup strategy for CSI-fed PMSM drives, ensuring a smooth transition to sensorless FOC in submersible pump applications. The proposed method addresses the challenges of accurately calculating the minimum $q$-axis current and correcting real-estimated frame alignment errors, offering a practical and reliable solution. Using an error compensation strategy, the transition from I-f startup to sensorless FOC is achieved seamlessly, progressively correcting the error between the estimated and real rotor positions. This results in minimal speed and current oscillations, improving system performance and stability. Both simulation and experimental results validate the strategy's effectiveness in reducing torque fluctuations and enhancing efficiency during critical transition stages. This work provides a robust solution for sensorless control in CSI-fed PMSM drives, supporting high reliability and smooth operation in industrial applications.